\documentclass[
]{ceurart}

\usepackage{listings}
\usepackage{natbib}
\usepackage{algorithmic}
\usepackage{textcomp}
\usepackage{multirow}
\usepackage{graphicx} 
\usepackage{subcaption} 
\usepackage{booktabs}
\usepackage{ragged2e}
\usepackage[normalem]{ulem}
\usepackage{xspace}
\usepackage{caption} 
\usepackage{array}
\usepackage{makecell}
\usepackage{bookmark}
\usepackage[table,xcdraw]{xcolor}
\usepackage{colortbl}
\usepackage{soul}
\usepackage{adjustbox}

\usepackage{xcolor}
\usepackage[most]{tcolorbox}
\usepackage{tabularx} 

\usetikzlibrary{positioning, shapes.geometric, calc}
\usetikzlibrary{arrows.meta, bending, shapes.misc}

\usepackage{tcolorbox}

\usepackage{xcolor}                 

\definecolor{specialblue}{HTML}{018786}

\definecolor{specialpurple}{HTML}{6A1B9A}

\usepackage{natbib}
\usepackage{algorithmic}
\usepackage{textcomp}
\usepackage{multirow}
\usepackage{graphicx} 
\usepackage{subcaption} 
\usepackage{booktabs}
\usepackage{ragged2e}
\usepackage[normalem]{ulem}
\usepackage{xspace}
\usepackage{caption} 
\usepackage{array}
\usepackage{makecell}
\usepackage{bookmark}
\usepackage[table,xcdraw]{xcolor}
\usepackage{colortbl}
\usepackage{soul}
\usepackage{adjustbox}

\usepackage{xcolor}
\usepackage[most]{tcolorbox}
\usepackage{tabularx} 

\usetikzlibrary{positioning, shapes.geometric, calc}
\usetikzlibrary{arrows.meta, bending, shapes.misc}

\usepackage{tcolorbox}

\usepackage{amsmath}
\usepackage{amssymb}
\usepackage{amsthm}
\usepackage{thmtools}
\usepackage{enumitem}

\newtheoremstyle{rqstyle}
  {0.0\baselineskip}{0.0\baselineskip}{\normalfont}{0pt}{\bfseries}{.}{0.0em}
  {\thmnumber{#2}\thmnote{ \normalfont(#3)}}
\theoremstyle{rqstyle}
\newtheorem{rqthm}{}
\renewcommand{\therqthm}{RQ\arabic{rqthm}}

\NewDocumentEnvironment{rqs}{o}
  {\IfValueTF{#1}{\rqthm[#1]}{\rqthm}\leavevmode}
  {\endrqthm}

\newlength{\rqlabelwidth}\newlength{\rqlabelsep}\newlength{\rqindent}
\newlist{subrqs}{enumerate}{1}
\setlist[subrqs]{
  label      = \textbf{\therqthm.\arabic*},
  ref        = \therqthm.\arabic*,
  align      = parleft,
  labelwidth = \rqlabelwidth,
  labelsep   = \rqlabelsep,
  leftmargin = \dimexpr\rqlabelwidth+\rqlabelsep+\rqindent\relax,
  itemindent = 0pt, itemsep = 0.4ex, topsep = 0.5ex, parsep = 0pt,
}

\declaretheoremstyle[
  spaceabove=6pt, spacebelow=6pt,
  headfont=\normalfont\bfseries,
  notefont=\mdseries, notebraces={(}{)},
  bodyfont=\normalfont,
  postheadspace=0.6em,
  headpunct=:
]{mystyle}

\usepackage[toc,acronyms,nonumberlist,nohypertypes={acronym}]{glossaries}
\newacronym{rq}{RQ}{research question}
\newacronym{ir}{IR}{Information Retrieval}
\newacronym{hci}{HCI}{Human-Computer Interaction}
\newacronym{genir}{GenIR}{Generative Information Retrieval}
\newacronym{ai}{AI}{Artificial Intelligence}
\newacronym{isar}{IS\&R}{Information Seeking and Retrieval}
\newacronym{neuro}{NeuroPhys}{neurophysiological}
\newacronym{sal}{SAL}{Search-as-Learning}
\newacronym{chi}{CHI}{Computer-Human Interaction}
\newacronym{llm}{LLM}{Large Language Model}
\newacronym{serp}{SERP}{search engine result page}
\newacronym{genai}{GenAI}{Generative AI}
\newacronym{rs}{RS}{Recommender System}
\newacronym{ui}{UI}{User Interface}
\newacronym{in}{IN}{information need}
\newacronym{mie}{MIE}{modern information environment}
\newacronym{isp}{ISP}{information seeking process}
\newacronym{is}{IS}{information seeking}
\newacronym{ar}{AR}{Augmented Reality}
\newacronym{sui}{SUI}{search user interface}
\newacronym{iim}{IIM}{Information Interaction Modality}
\newacronym{rag}{RAG}{Retrieval-Augmented Generation}
\newacronym{seo}{SEO}{search engine optimization}
\newacronym{cit}{CIT}{Critical Incident Technique}
\newacronym{dce}{DCE}{Discrete Choice Experiment}
\newacronym{bws}{BWS}{Best-Worst Scaling}
\newacronym{bwdce}{BWDCE}{Best-Worst Discrete Choice Experiment}
\newacronym{opuf}{OPUF}{Online Elicitation of Personal Utility Functions}
\newacronym{mmnl}{MMNL}{Mixed Multinomial Logit Model}
\newacronym{eeg}{EEG}{Electroencephalography}
\newacronym{eda}{EDA}{Electrodermal Activity}
\newacronym{ddtf}{dDTF}{dynamic Direct Transfer Function} 

\usepackage{cleveref}

\crefformat{rqthm}{#2#1#3}\Crefformat{rqthm}{#2#1#3}
\crefformat{subrqsi}{#2#1#3}\Crefformat{subrqsi}{#2#1#3}
\crefformat{hyp}{#2#1#3}\Crefformat{hyp}{#2#1#3}
\crefrangeformat{rqthm}{#3#1#4 to #5#2#6}\Crefrangeformat{rqthm}{#3#1#4 to #5#2#6}
\crefrangeformat{subrqsi}{#3#1#4 to #5#2#6}\Crefrangeformat{subrqsi}{#3#1#4 to #5#2#6}
\crefrangeformat{hyp}{#3#1#4 to #5#2#6}\Crefrangeformat{hyp}{#3#1#4 to #5#2#6}
\crefmultiformat{rqthm}{#2#1#3}{ and #2#1#3}{, #2#1#3}{ and #2#1#3}
\crefmultiformat{subrqsi}{#2#1#3}{ and #2#1#3}{, #2#1#3}{ and #2#1#3}
\crefmultiformat{hyp}{#2#1#3}{ and #2#1#3}{, #2#1#3}{ and #2#1#3}

\begin{document}

\copyrightyear{2026}
\copyrightclause{Copyright for this paper by its authors.
  Use permitted under Creative Commons License Attribution 4.0
  International (CC BY 4.0).}

\conference{14th PhD Symposium on Future Directions in Information Access (FDIA 2026), July 30, 2026, Bucharest, Romania}

\title{{Characterizing the Evolving Landscape of\\Modern Information Seeking}}


\author[1]{Shuoqi Sun}[%
orcid=0009-0000-9329-9731,
email=shuoqi.sun@student.rmit.edu.au,
url=https://shuoqisun.github.io/,
]
\address[1]{RMIT University, Melbourne, Australia}




\begin{abstract}
Information seeking (IS) evolves, as does the human IS process. Since the rise of Generative AI (GenAI), modern IS has shifted by introducing more interfaces, more complex interactions, and expanded system capabilities. We argue that these changes in modern IS should be systematically examined. This PhD research characterizes the changes in the modern IS process. We use mechanisms, including online crowdsourcing survey experiments, theoretical IS frameworks, and in-lab experiments with neurophysiological signals, to characterize the shifts in modern IS, especially those driven by GenAI. We offer insights into the current landscape of search interface preferences and the cognitive efforts involved in seeking information. We believe this PhD research will contribute to and inform future designs of personalized, cognition-aware IS systems.
\end{abstract}


\maketitle

\vspace{-4mm}

\section{Introduction}
\vspace{-1mm}

Information environments evolve, and emerging information technologies can reshape human behavior and cognition~\citep{ISMIE, Sparrow2011Science_Google_Effect}. In recent years, \gls{genai} chatbots have gained popularity and reshaped the \gls{isp}~\citep{White2025Book_IA_GenAI}. Within \glspl{mie}, \glspl{sui} have diversified, and \gls{genai} chatbots intensify this trend. We argue that the landscape has shifted, and its changes require systematic attention. Understanding what people adopt, and how, is critical to informing future \gls{ir} system design and interaction paradigms. This PhD research investigates how the diversification of \glspl{sui}, especially \gls{genai} chatbots, reshapes \gls{is} in terms of \gls{sui} preference, \gls{isp} constructs, and the human cognition involved. We believe this PhD research will contribute to future designs of personalized~\citep{shah2025fromtodototada}, cognition-aware~\citep{Moshfeghi2025SIGIR_BMI} \gls{ir} systems and ecosystems.

\vspace{-3mm} 
\section{Proposed Research Questions and Methodology}
\label{sec:rqs-methods}
\vspace{-1mm}

Our \glspl{rq} are listed below in the conceptual order of this PhD research:

\begin{rqs}
\label{rq:1}
Do we have a conceptual framework to characterize modern \gls{is}? If not, what would one look like?
\end{rqs}

\begin{rqs}
\label{rq:2}
How do people choose among diversified \glspl{sui} according to search scenarios within \glspl{mie}, and which contextual and interface factors drive their preferences?
\end{rqs}

\begin{rqs}
\label{rq:3}
Within modern \glspl{isp}, how do people's cognitive efforts differ when interacting with information produced by traditional search engines versus \gls{genai} chatbots? Specifically, if information coverage is controlled, which presentation factors affect human cognition?
\end{rqs}

\begin{rqs}
\label{rq:4}
(Preliminary) Can we characterize the \glspl{isp} with \gls{genai} chatbots using \gls{neuro} signals, emphasizing multi-turn interactions? How do the neuro correlates differ in \glspl{isp} with \gls{genai} chatbots than with search engines?

\end{rqs}


\noindent
We examine how the rise of emerging \glspl{sui} (including \gls{genai} chatbots) affect modern \gls{isp}. We first raise a fundamental question (\cref{rq:1}): does a framework exist to theoretically characterize the evolved \gls{isp} within \gls{mie}? Theoretical frameworks are critical conceptual instruments that can be used to characterize information environments and identify confounding variables. To address \cref{rq:1}, we conduct a literature review, which reveals a conceptual gap within \gls{mie}. We therefore propose the Information Seeking in Modern Information Environments framework (ISMIE)~\citep{ISMIE}, which conceptualizes and characterizes the modern \gls{isp}. The ISMIE framework identifies four essential components, four activities and six variables, emphasizing the dynamics within \gls{mie}.

We then view the larger picture of people's \gls{is} behavior (\cref{rq:1}), characterizing it by modeling human preferences for primary \glspl{sui}, including search engines, \gls{genai} chatbots, and social media platforms. 
We conduct a two-phase online crowdsourcing survey study to involve a larger population. We believe contexts are critical. Therefore, we first collect real-life search scenarios via structured survey questions in Phase~1. The collected scenarios are manually reviewed and standardized into a uniform template. In Phase~2, the context-dependent discrete choice experiments~\citep{vanKasteren2026UMAP_Preference, lancsar2013_BWDCE} are conducted to understand people's preferences for \glspl{sui}. We present templated scenarios and ask participants to indicate their preferences among \gls{sui} alternatives. Then by using qualitative and quantitative analyses, we surface the contextual and interface factors that shape people's preferences.

We anticipate a large use of \gls{genai} chatbots. However, evidence shows that the use of \gls{genai} chatbots cause cognitive effects~\cite{ Kosmyna2025arXiv_Cognitive_Debt, Stadler2024_Cognitive_Ease, Shukla2025CHIEA_Ironies, Sterling2026Science_AI_Attitude_Bias} and potentially cognitive decline~\cite{Noorbehbahani2026H1_Cognitive_Decline}. Such a fact prompts \gls{ir} researchers to investigate and contribute from \gls{is} perspectives. \gls{ir} community has paid attention on user cognition and studied \gls{isp} through brain activity~\citep[e.g.,][]{Moshfeghi2025SIGIR_BMI, Ji2024SIGIR_Physiological, Moshfeghi2018WWW_Search, NeuroPhysIIR2025CHIIR,spina2025report}, offering empirical practices and evidence for the feasibility of deeper investigation. To address \cref{rq:3}, we conduct in-lab experiments and use \gls{neuro} signals (e.g., \gls{eeg}) to characterize the effects of information produced by traditional search engines and \gls{genai} chatbots.
Specifically, we ask a question: \emph{if information coverage is the same, then which factors affect user cognition?} We focus on the \emph{acquiring} activity in the ISMIE framework where people examine the presented information and alter their internal cognitive states. Our experiments focus on two aspects, potentially trade-offs: cognitive effort and search outcomes. Cognitive effort, such as cognitive load and attention, is represented through multiple \gls{neuro} signals~\cite{Ji2024SIGIR_Physiological}. In experiments, participants are required to execute search tasks and produce search outcomes based on provided information, during which their \gls{neuro} signals are recorded. The provided information is then systematically altered to answer \cref{rq:3}.


The \cref{rq:4} is preliminary and proposed to extend~\citet{Moshfeghi2018WWW_Search} and~\citet{Ji2024SIGIR_Physiological}'s understanding of \gls{isp}. As a further investigation, we emphasize the multi-modal and multi-turn interactions that are common within \gls{mie}~\cite{ISMIE}. Our experiments mimic the interactions between users and \gls{genai} chatbots. This study is necessary due to shifts in the interaction paradigm, such as from keyword queries to prompts, follow-up questions in a single search session, etc.




\vspace{-3mm}
\section{Attending FDIA 2026 and Future Plans}
\vspace{-1mm}

This PhD research is positioned at the intersection of \gls{ir}, \gls{hci}, and cognitive science. 
The key components of this PhD research
intersect with the expertise of the FDIA community. I anticipate meaningful discussion and feedback from FDIA 2026 attendees.

The ISMIE framework for \cref{rq:1} was published at SIGIR-AP'25~\cite{ISMIE}, serving as a cornerstone of this PhD research.
The work for \cref{rq:2} is granted ethics approval (project ID: 30284) and is now ready to proceed with experiments
The investigations of \cref{rq:3} and \cref{rq:4} have a preliminary methodology design that will benefit substantially from discussion at FDIA 2026.
I believe FDIA 2026 will help inspire their methodology designs, positioning, and potential collaborations with like-minded researchers.








\vspace{-3mm}
\begin{acknowledgments}
\vspace{-1mm}

The author thanks reviewers for their constructive feedback and inspiring words. The author thanks his PhD supervisors, Dr.~Damiano Spina and Dr.~Danula Hettiachchi, for their constant guidance and support. The author thanks the ARC Center of Excellence for Automated Decision-Making and Society (ADM+S) for funding his PhD research. The author also thanks ADM+S and the European Summer School on Information Retrieval (ESSIR 2026) organizing committee for their travel grant. The author pays his respect to the Woi wurrung and Boon wurrung language groups of the eastern Kulin Nation, on whose unceded lands this research is conducted, and Ancestors and Elders, past and present.
\end{acknowledgments}

  


\clearpage
\bibliography{99_ref}

@article{spina2025report,
author = {Spina, Damiano and Gwizdka, Jacek and Ji, Kaixin and Moshfeghi, Yashar and Mostafa, Javed and Ruotsalo, Tuukka and Zhang, Min and Ahmad, Adnan and Lawati, Sara Fahad Dawood Al and Boonprakong, Nattapat and Fernando, Nishani and He, Jiaman and Hoeber, Orland and Jayawardena, Gavindya and Lee, Boon-Giin and Liu, Haiming and Pike, Matthew and Pirmoradi, Abbas and Nakisa, Bahareh and Rastgoo, Mohammad Naim and Salim, Flora D. and Scott, Fletcher and Sun, Shuoqi and Tang, Huimin and Towey, Dave and Wilson, Max L.}, title = {{Report on the 3rd Workshop on NeuroPhysiological Approaches for Interactive Information Retrieval (NeuroPhysIIR 2025) at SIGIR CHIIR 2025}}, year = {2025}, issue_date = {June 2025}, publisher = {Association for Computing Machinery}, address = {New York, NY, USA}, volume = {59}, number = {1}, issn = {0163-5840}, doi = {10.1145/3769733.3769740}, journal = {SIGIR Forum}, month = oct, pages = {1–43}, numpages = {43} }

@inproceedings{shah2025fromtodototada,
  author    = {Shah, Chirag and White, Ryen W.},
  title     = {{From To-Do to Ta-Da: Transforming Task-Focused IR with Generative AI}},
  year      = {2025},
  isbn      = {9798400715921},
  publisher = {Association for Computing Machinery},
  address   = {New York, NY, USA},
  doi       = {10.1145/3726302.3730352},
  booktitle = {Proceedings of the 48th International ACM SIGIR Conference on Research and Development in Information Retrieval},
  pages     = {3911--3921},
  numpages  = {11},
  location  = {Padua, Italy},
  series    = {SIGIR '25}
}

@article{lancsar2013_BWDCE,
  title   = {{Best Worst Discrete Choice Experiments in Health: Methods and an Application}},
  journal = {Social Science \& Medicine},
  volume  = {76},
  pages   = {74--82},
  year    = {2013},
  issn    = {0277-9536},
  doi     = {10.1016/j.socscimed.2012.10.007},
  author  = {Emily Lancsar and Jordan Louviere and Cam Donaldson and Gillian Currie and Leonie Burgess}
}

@inproceedings{vanKasteren2026UMAP_Preference,
author = {van Kasteren, Anouk and Vredenborg, Marloes and Bauer, Christine and Masthoff, Judith},
title = {{Modelling Preference Heterogeneity for Context-Aware Decision Support During Public Transport Disruptions}},
year = {2026},
isbn = {9798400723117},
publisher = {Association for Computing Machinery},
address = {New York, NY, USA},
doi = {10.1145/3774935.3806191},
booktitle = {Proceedings of the 34th ACM Conference on User Modeling, Adaptation and Personalization},
pages = {156–165},
numpages = {10},
location = {
},
series = {UMAP '26}
}

@inproceedings{Moshfeghi2025SIGIR_BMI,
author = {Moshfeghi, Yashar and Mcguire, Niall},
title = {{Brain-Machine Interfaces \& Information Retrieval Challenges and Opportunities}},
year = {2025},
isbn = {9798400715921},
publisher = {Association for Computing Machinery},
address = {New York, NY, USA},
doi = {10.1145/3726302.3730350},
booktitle = {Proceedings of the 48th International ACM SIGIR Conference on Research and Development in Information Retrieval},
pages = {3887–3898},
numpages = {12},
location = {Padua, Italy},
series = {SIGIR '25}
}

@inproceedings{ISMIE,
author = {Sun, Shuoqi and Hettiachchi, Danula and Spina, Damiano},
title = {{ISMIE: A Framework to Characterize Information Seeking in Modern Information Environments}},
year = {2025},
isbn = {9798400722189},
publisher = {Association for Computing Machinery},
address = {New York, NY, USA},
doi = {10.1145/3767695.3769509},
booktitle = {Proceedings of the 2025 Annual International ACM SIGIR Conference on Research and Development in Information Retrieval in the Asia Pacific Region},
pages = {385–395},
numpages = {11},
location = {China},
series = {SIGIR-AP 2025}
}

@article{Sparrow2011Science_Google_Effect,
author = {Betsy Sparrow  and Jenny Liu  and Daniel M. Wegner },
title = {{Google Effects on Memory: Cognitive Consequences of Having Information at Our Fingertips}},
journal = {Science},
volume = {333},
number = {6043},
pages = {776-778},
year = {2011},
doi = {10.1126/science.1207745},
}

@inproceedings{Ji2024SIGIR_Physiological,
author = {Ji, Kaixin and Hettiachchi, Danula and Salim, Flora D. and Scholer, Falk and Spina, Damiano},
title = {{Characterizing Information Seeking Processes with Multiple Physiological Signals}},
year = {2024},
isbn = {9798400704314},
publisher = {Association for Computing Machinery},
address = {New York, NY, USA},
doi = {10.1145/3626772.3657793},
booktitle = {Proceedings of the 47th International ACM SIGIR Conference on Research and Development in Information Retrieval},
pages = {1006–1017},
numpages = {12},
location = {Washington DC, USA},
series = {SIGIR '24}
}

@inproceedings{Moshfeghi2018WWW_Search,
author = {Moshfeghi, Yashar and Pollick, Frank E.},
title = {{Search Process as Transitions Between Neural States}},
year = {2018},
isbn = {9781450356398},
publisher = {International World Wide Web Conferences Steering Committee},
address = {Republic and Canton of Geneva, CHE},
doi = {10.1145/3178876.3186080},
booktitle = {Proceedings of the 2018 World Wide Web Conference},
pages = {1683–1692},
numpages = {10},
location = {Lyon, France},
series = {WWW '18}
}

@article{Noorbehbahani2026H1_Cognitive_Decline,
title = {{AI-Overdependence and Human Cognitive Decline: Hazards, Evidence, and Mitigation Strategies}},
journal = {Computers in Human Behavior Reports},
volume = {22},
pages = {101102},
year = {2026},
issn = {2451-9588},
doi = {10.1016/j.chbr.2026.101102},
author = {Fakhroddin Noorbehbahani and Kiemute Oyibo},
}

@inproceedings{Shukla2025CHIEA_Ironies,
author = {Shukla, Prakash and Bui, Phuong and Levy, Sean S and Kowalski, Max and Baigelenov, Ali and Parsons, Paul},
title = {{De-skilling, Cognitive Offloading, and Misplaced Responsibilities: Potential Ironies of AI-Assisted Design}},
year = {2025},
isbn = {9798400713958},
publisher = {Association for Computing Machinery},
address = {New York, NY, USA},
doi = {10.1145/3706599.3719931},
booktitle = {Proceedings of the Extended Abstracts of the CHI Conference on Human Factors in Computing Systems},
articleno = {171},
numpages = {7},
}

@misc{Kosmyna2025arXiv_Cognitive_Debt,
  title         = {{Your Brain on ChatGPT: Accumulation of Cognitive Debt when Using an AI Assistant for Essay Writing Task}},
  author        = {Nataliya Kosmyna and Eugene Hauptmann and Ye Tong Yuan and Jessica Situ and Xian-Hao Liao and Ashly Vivian Beresnitzky and Iris Braunstein and Pattie Maes},
  year          = {2025},
  archiveprefix = {arXiv},
  primaryclass  = {cs.AI},
  url           = {https://arxiv.org/abs/2506.08872}
}

@article{Stadler2024_Cognitive_Ease,
  title   = {{Cognitive Ease at a Cost: LLMs Reduce Mental Effort but Compromise Depth in Student Scientific Inquiry}},
  journal = {Computers in Human Behavior},
  volume  = {160},
  pages   = {108386},
  year    = {2024},
  issn    = {0747-5632},
  doi     = {https://doi.org/10.1016/j.chb.2024.108386},
  author  = {Matthias Stadler and Maria Bannert and Michael Sailer}
}

@article{Sterling2026Science_AI_Attitude_Bias,
  author  = {Sterling Williams-Ceci  and Maurice Jakesch  and Advait Bhat  and Kowe Kadoma  and Lior Zalmanson  and Mor Naaman },
  title   = {{Biased AI Writing Assistants Shift Users’ Attitudes on Societal Issues}},
  journal = {Science Advances},
  volume  = {12},
  number  = {11},
  pages   = {eadw5578},
  year    = {2026},
  doi     = {10.1126/sciadv.adw5578}
}

@book{White2025Book_IA_GenAI,
  title={Information Access in the Era of Generative AI},
  editor={White, Ryen W. and Shah, Chirag},
  year={2025},
  publisher={Springer Cham},
  doi={10.1007/978-3-031-73147-1}
}

@inproceedings{NeuroPhysIIR2025CHIIR,
author = {Gwizdka, Jacek and Mostafa, Javed and Zhang, Min and Ji, Kaixin and Moshfeghi, Yashar and Ruotsalo, Tuukka and Spina, Damiano},
title = {{NeuroPhysIIR: International Workshop on NeuroPhysiological Approaches for Interactive Information Retrieval}},
year = {2025},
isbn = {9798400712906},
publisher = {Association for Computing Machinery},
address = {New York, NY, USA},
doi = {10.1145/3698204.3716481},
booktitle = {Proceedings of the 2025 ACM SIGIR Conference on Human Information Interaction and Retrieval},
pages = {413–415},
numpages = {3},
location = {
},
series = {CHIIR '25}
}




\end{document}